\documentclass[letterpaper]{article}
\usepackage{iccc}

\usepackage{times}
\usepackage{helvet}
\usepackage{courier}

\usepackage{graphicx}
\usepackage{subcaption}
\usepackage{xcolor}

\definecolor{transcription}{RGB}{0,150,0}
\definecolor{omr}{RGB}{0,0,200}
\definecolor{lm_m}{RGB}{150,50,0}

\mathchardef\mhyphen="2D

\title{Language Model Mapping in Multimodal Music Learning: \\
A Grand Challenge Proposal}

\author{
    Daniel Chin$^{1,2}$ \quad Gus Xia$^{2}$ \\
    $^1$NYU Shanghai \quad $^2$MBZUAI \\
    \texttt{daniel.chin@nyu.edu} \quad \texttt{gus.xia@mbzuai.ac.ae}
}

\begin{document} 
\maketitle
\begin{abstract}
\begin{quote}
We have seen remarkable success in representation learning and language models (LMs) using deep neural networks. Many studies aim to build the underlying connections among different modalities via the alignment and mappings at the token or embedding level, but so far, most methods are very data-hungry, limiting their performance in domains such as music where paired data are less abundant. We argue that the embedding alignment is only at the surface level of multimodal alignment. In this paper, we propose a grand challenge of \textit{language model mapping} (LMM), i.e., how to map the essence implied in the LM of one domain to the LM of another domain under the assumption that LMs of different modalities are tracking the same underlying phenomena. We first introduce a basic setup of LMM, highlighting the goal to unveil a deeper aspect of cross-modal alignment as well as to achieve more sample-efficiency learning. We then discuss why music is an ideal domain in which to conduct LMM research. After that, we connect LMM in music with a more general and challenging scientific problem of \textit{learning to take actions based on both sensory input and abstract symbols}, and in the end, present an advanced version of the challenge problem setup.
\end{quote}
\end{abstract}

\section{Introduction}
Multimodality refers to the use of multiple sensory channels, such as visual, aural, and gestural channels, in processing and communicating information. Different modalities provide diverse perspectives or portrayals of the same objective world. From a time-series perspective, though representations in different modalities may seem to follow distinct rules, such as those prescribed by language models (LMs), those models fundamentally track the moment-to-moment evolution of the same underlying phenomena. This concept forms the basis of what is known as \textit{cross-modal alignment}.


Current machine-learning studies on multimodality are primarily built on the alignment and mapping between multimodal tokens or their embeddings extracted from short segments. Furthermore, the alignments are usually trained by a large quantity of paired data. For example, CLIP \cite{Radford2021clip} adopts contrastive learning to map images and short textual captions to a joint latent space, which serves as a foundation for many follow-up studies on image understanding and generation \cite{koh2024GILL,yu2024spae,zheng2023minigpt5,wu2023NExT-GPT}. Similarly, CLAP \cite{wu2023clap} learns to align audio clips and text descriptions, and in a recent study, CLaMP \cite{wu2023clamp} aligns symbolic music score with text descriptions about the music. 

\begin{figure*}[t]
    \centering

    \begin{subfigure}[]{0.45\textwidth}
        \centering
        \includegraphics[page=3,scale=0.4]{figs/triangle.pdf}
        \caption{}
    \end{subfigure}
    ~
    \begin{subfigure}[]{0.45\textwidth}
        \centering
        \includegraphics[scale=0.4]{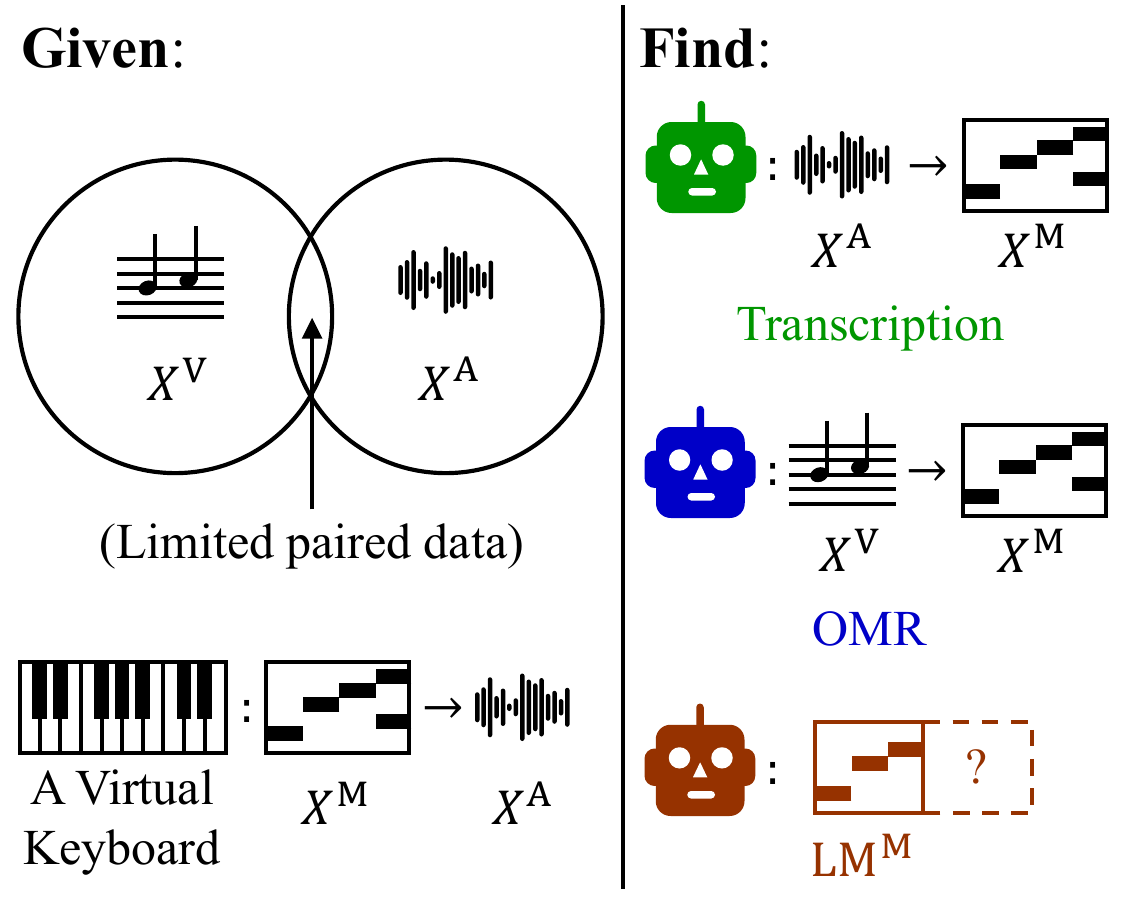}
        \caption{}
    \end{subfigure}
    
    \caption{(a) The three modalities in music. Vision (V) refers to music score images. Audio (A) refers to music audio. Motion (M) refers to instrument controls, i.e., detailed performance motions. Data in each modality can be modeled by a uni-modal LM. Arrows across modalities refer to time-aligned translation tasks, including OMR and transcription. (b) Illustration of the basic setup. The basic version of our grand challenge involves three given elements and three desired goals. Data are abundantly available in A and V, and the synthesis task is considered to be already well-solved by rule-based music synthesizers as virtual instruments. Given the above, we seek the three goals marked with unique colors: OMR, transcription, and an LM in M.}
    \label{fig:basic-setup}
\end{figure*}

We argue that human learning is obviously more sample-efficient, and the information flow of different modalities connects more deeply in human minds than in current machine-learning models. For example, humans never need to learn from millions of paired video clips and captions to describe videos. Rather, we learn how visual representations of the world flow by \textit{observing}, and we learn how to write proper texts by \textit{reading}. These two processes need almost no paired data but they help each other --- when reading, the mind imagines the scenes; when watching a movie, the mind narrates what is happening. Based on the cross-modal alignment assumption that these two information flows eventually describe the same underlying world, humans only need a small amount of paired data to calibrate the already developed mapping. 

The deep connections between different modalities and the resulting sample-efficient benefits are more apparent when we further consider an \textit{action} model. For example, when learning to improvise on the piano, i.e., to build an action language model, the music we have read and listened to plays a significant role. Musicians learn internal language models for audio and symbolic music, and the musicality implied in the language models can be mapped to the process of piano playing, freeing us from rebuilding the musicality from scratch by literally performing abundant pieces. Additionally, based on the language models, a pianist can learn to sight read a score and to transcribe piano performance from audio with no paired data but purely by trial-and-error on a keyboard.

Inspired by the aforementioned observations, we believe \textit{the alignment of tokens and embeddings is only at the surface level of multimodal representation learning}. In this paper, we propose a grand challenge of \textbf{Language model mapping (LMM)} between different modalities, in order to unveil a deeper aspect of cross-modal alignment and the efficiency of human learning.

Music is an ideal domain for LMM research for three reasons. Firstly, music data naturally involve multiple closely aligned modalities that seem superficially distinct: music audio, score, instrument control (MIDI), etc. Secondly, compared to other domains such as robot control and scene rendering, the virtual environment of the music world is concise and clean, as the control of the instruments is low-dimensional and the consequences of actions are simple and locally contained. Lastly, the sparsity of music data compared to vision and text domains forces music AI researchers to develop more sample-efficient solutions. 

Fig.\ref{fig:basic-setup}(a) shows the framework of LMM in music. A, audio, refers to the modality of music audio. V, vision, refers to the modality of score images. M, motion, refers to the modality of instrument controls, i.e., the performance motions. Data within each modality can have their distribution modeled, yielding $\mathrm{LM}^\mathrm{A}$, $\mathrm{LM}^\mathrm{V}$, and $\mathrm{LM}^\mathrm{M}$. Across the modalities are translation tasks including optical music recognition (OMR) and music transcription. The synthesis from H to A is usually considered well-solved by rule-based virtual instruments. The challenge of LMM in music, with a basic example shown in Fig.\ref{fig:basic-setup}(b), thus lies in learning LMs or cross-modal maps utilizing not only within-modal data but also data and models from other modalities.

\section{Grand Challenge Statement}
This section defines the grand challenge of Language Model Mapping (LMM) in Multimodal Music Learning by presenting a basic problem setup, contrasting related music AI tasks, and discussing how humans leverage LMM to achieve efficient multimodal learning.

\subsection{Problem Setup: Basic Version}
\label{subsec:problem-setup-basic}
The basic version of our grand challenge poses a simple question: Can we learn to play a given instrument by listening to a lot of music and reading a lot of scores? Fig.\ref{fig:basic-setup} shows the problem setup. The following elements are considered \textbf{given}: 
\begin{itemize}
    \item an abundance of score image data, \\
    $X^\mathrm{V} = \{ \mathbf{x}_i^\mathrm{V} \}_{i=1}^n,$ where $\mathbf{x}_i^\mathrm{V} \in \mathcal{X}^\mathrm{V}$,
    \item an abundance of music audio data, \\
    $X^\mathrm{A} = \{ \mathbf{x}_i^\mathrm{A} \}_{i=1}^m,$ where $\mathbf{x}_i^\mathrm{A} \in \mathcal{X}^\mathrm{A},$
    \item a virtual differentiable piano keyboard capable of synthesizing audio from instrument controls, \\
    $f: \mathcal{X}^\mathrm{M} \to \mathcal{X}^\mathrm{A},$
    \item and a limited supply of paired score-audio data, \\
    $S^\mathrm{V \mhyphen A} = \{ ( \mathbf{x}_i^\mathrm{V}, \mathbf{x}_i^\mathrm{A} ) \}_{i=1}^l \subset \mathcal{X}^\mathrm{V} \times \mathcal{X}^\mathrm{A},$ \\
    where $l \ll n, l \ll m.$ 
\end{itemize}
With the above elements available, \textbf{obtain} the following: 
\begin{itemize}
    \item \textcolor{omr}{
    the capability of OMR from score to performance motions, \\
    $g: \mathcal{X}^\mathrm{V} \to \mathcal{X}^\mathrm{M},$
    }
    \item \textcolor{transcription}{
    the capability of transcription from audio to performance motions, \\
    $h: \mathcal{X}^\mathrm{A} \to \mathcal{X}^\mathrm{M},$
    }
    \item \textcolor{lm_m}{
    and a language model in performance motions, \\
    $\mathrm{LM}^\mathrm{M}.$ 
    }
\end{itemize}

\subsection{Related Music AI Problems}
Several existing learning tasks in computer music are related to our LMM grand challenge setup. In the self-supervised language modeling of music signal or tokens, a uni-modal setup is usually implied (with a few recent exceptions discussed in the Existing Approaches Section). In the OMR and music transcription tasks, the setup usually does not involve any language modeling, so it almost always requires supervision by paired data. Unsupervised OMR and music transcription is usually impossible without strong domain knowledge. The grand challenge of LMM opposes the dichotomy of supervised v.s. unsupervised which only makes sense in uni-modal setups. The grand challenge of LMM essentially invites researchers to open up the modality landscape, recognize the holistic closed loop of human's multimodal musicality, and design a reasonable model pipeline that utilizes LMs and data in related modalities as well as cross-modal mappings.

\subsection{Baseline Solution v.s. Human Approach}
A naive baseline approach to the simple version of the challenge problem is to first somehow translate the available audio and score data to the performance motion modality and then train an $\mathrm{LM}^\mathrm{M}$ on the translated data. However, that is clearly not how humans learn to play music. In fact, the above proposed translation tasks are ill-formed, because A contains more information than M while V contains not enough information to specify M. Translating everything to M is not a grounded approach. In contrast, humans use closed loops to learn skills and concepts. From V and A data humans construct $\mathrm{LM}^\mathrm{V}$ and $\mathrm{LM}^\mathrm{A}$. Both LMs model \textit{musicality} from a modality-specific angle. From the internalization of musicality, humans somehow learn to play the musical instrument via LMM. The challenge for the AI community is to study exactly how LMM prescribes and facilitates that learning process. Such is the research question of LMM in Multimodal Music Learning. 

\section{Existing Approaches to Multimodal Language Modeling}
\label{sec:related-work}
NExT-GPT \cite{wu2023NExT-GPT}, GILL \cite{koh2024GILL}, Mini-GPT5 \cite{zheng2023minigpt5}, and AudioLDM 2 \cite{liu2023audioldm2} each augments a pretrained large natural language model to perceive and generate multimodal data. They all insert special tokens into the natural language token sequence to refer to cross-modal information. Those methods show impressive end-to-end multimodal processing capabilities, and one may even conclude that those LLM agents effectively understands the semantics of the cross-modal content referenced by those special tokens. Importantly, their methods work well because their augmented natural language still follows the same general patterns as the vanilla natural language: suppose the human natural language vocabulary was expanded to include rich ``pronouns'' that refer to images and sounds, then using those pronouns in regular sentences would still generally follow the form, grammar, and patterns of the vanilla natural language. 

However, the same logic does not hold for music understanding and composition. A musical piece, when broken down to a sequence of references to music elements or music effects, does not follow the patterns of natural language. The form and grammar of music and natural language differ so much that the approach of embedding music elements as text tokens for text-pretrained sequence modeling is, to say the least, remote from how musicality works as we humans understand it. As a result, any music-to-music process model-able by a text LM can at most be on the metadata level and never be on the musicality level. Any musical semantics beyond textual grounding will be lost by the text LM, prohibiting proper music modeling.

Using the natural language modality as the intermediary or central backbone for multimodal processing is an appealing and profound methodology, but it is limited because textual grounding alone cannot always cover the vast multimodal grounding available to human learning. The grand challenge of Language Model Mapping does not assume or force any single backbone modality, but instead emphasizes both the unique nature of each modality and the alignment and mapping between those modalities, calling for more diverse approaches than existing studies.

\begin{figure}[t]
    \centering
    \includegraphics[scale=0.4]{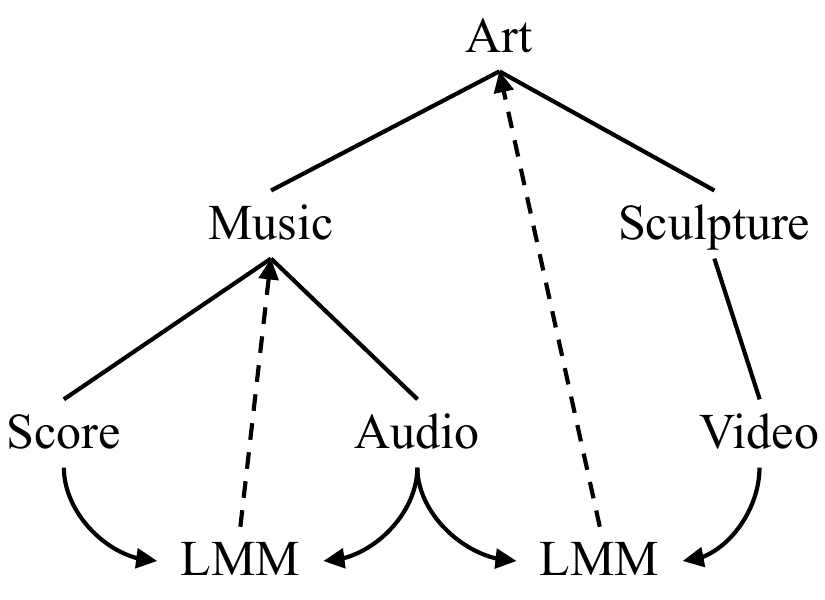}
    \caption{An example that LMM can be cross-domain. The within-domain LMM between music score and audio is contingent on different modalities sharing the same musicality. The cross-domain LMM between music audio and sculpture video is contingent on different domains of art sharing the same artistic nature.}
    \label{fig:cross-domain}
\end{figure}

\section{Significance of LMM from Perspective of AI}
\subsection{Learning Actions from Both Abstract Symbols and Raw Sensory Input}
There is the general challenge of learning actions from just abstract symbols and raw sensory input, of which LMM serves as a specific case. LMM applies not only to the domain of music, but is instead generally present in many cognitive processes at large. Here are some examples. 

\begin{enumerate}
    \item Decipher a novel language and learn to speak it. Given texts and speech audio in a novel language and domain knowledge about speech synthesis, obtain \textcolor{transcription}{a speech-to-motion transcriber,} \textcolor{omr}{a text-to-motion pronouncer,} \textcolor{lm_m}{and most importantly a motion language model,} where ``motion'' refers to the time series of speech synthesis parameters. 
    \item Learn to cook, using only cookbooks and tutorial videos. In a robot-kitchen setup, given cooking videos, text cookbooks, and the capability to view how the scene reacts to robot motions, obtain \textcolor{transcription}{a video-following robot cook,} \textcolor{omr}{a cookbook-following robot cook,} \textcolor{lm_m}{and most importantly a cooking motion language model.} 
    \item Learn to interact with the world from reading and observation. Given natural language data, natural video data, and the capability to view how the scene reacts to robot motions, obtain \textcolor{transcription}{a video-following robot,} \textcolor{omr}{a text-following robot,} \textcolor{lm_m}{and most importantly an action language model, based on which the robot can interact with the world in general.} 
\end{enumerate}

Those examples show that LMM is applicable to a wide variety of domains. Compared to those complex setups, the music learning setup is simple and therefore presents a suitable first challenge, meanwhile closely following the formulation of LMM. 

\subsection{Learning Efficiency}
Joint embedding methods such as CLIP \cite{Radford2021clip} and CLAP \cite{wu2023clap} require a large amount of paired data. However, an ideal, human-like LM should be efficient. LMM promises learning efficiency on three levels. Firstly, it seeks to train a language model on M even when data in M is scarce by leveraging data from other modalities. Secondly, it trains a transcription and an OMR module without readily paired data leveraging multimodal LMs. Thirdly, in the ideal from of LMM, instead of constructing a separate uni-modal LM for each modality, we use multimodal joint training or training a pan-modal core model. Either way, LMM avoids repetitive work by unifying modalities under the assumption that different modalities only describe the one same musicality. In conclusion, LMM is both sample-efficient and training-efficient.

\subsection{LMM can be Cross-Domain}

In addition to being cross-modal, LMM can be cross-domain. As illustrated by the example in Fig.\ref{fig:cross-domain}, there can be higher levels of alignment between LMs. LMM between music score and music audio works on the assumption that the score and the audio share the same underlying musicality. Going one level up, LMM between music audio and sculpture video may work on the assumption that music and sculpture share something deeper in common. The far-end goal is to unify all modalities and domains. Indeed, from the perspective of human learning, it is single-task learning that stands out as an anomaly.

\section{Problem Setup: Advanced Version}

\begin{figure}[t]
    \centering
    \includegraphics[page=2,scale=0.4]{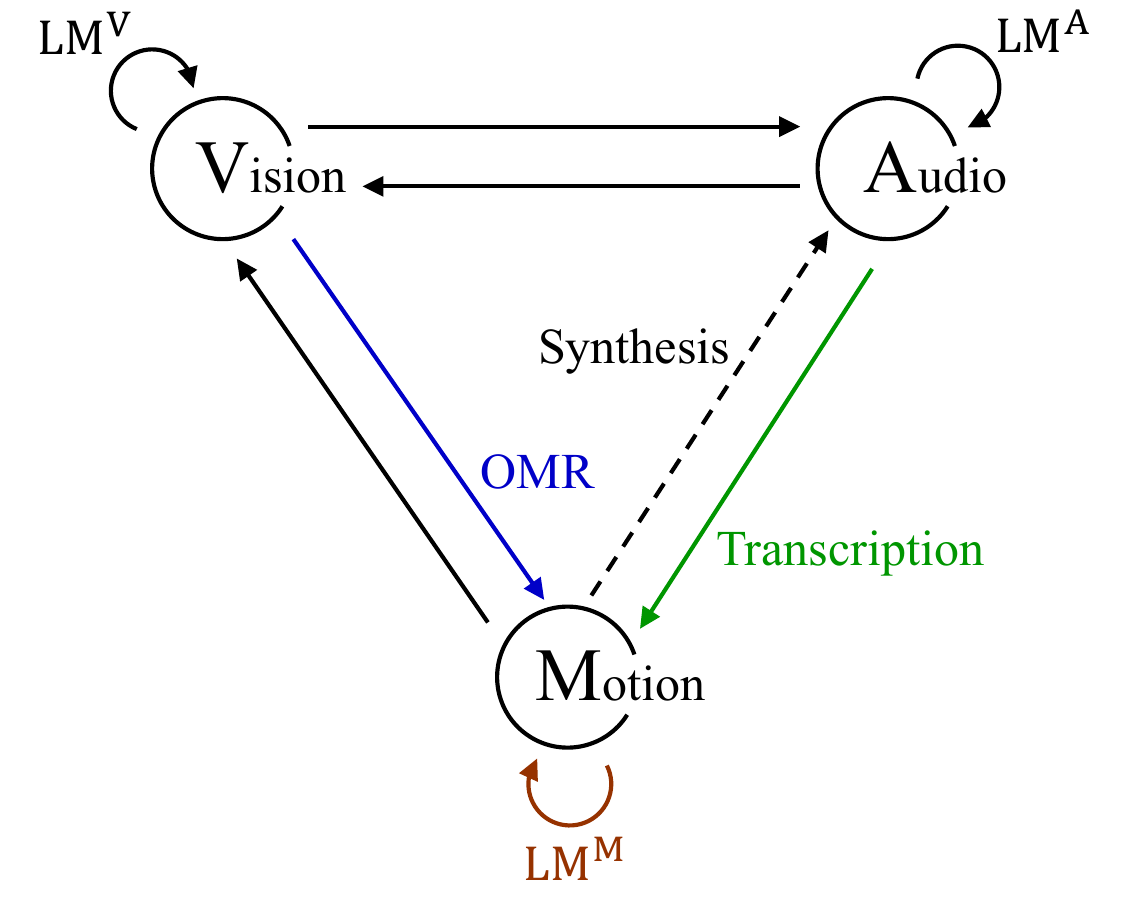}
    \caption{The advanced version of Fig.\ref{fig:basic-setup}(a), generalizing the LMM challenge in multimodal music learning. Vision (V) refers to music score images. Audio (A) refers to music audio. Motion (M) refers to instrument controls, i.e., detailed performance motions. Data in each modality can be modeled by a uni-modal LM. Arrows across modalities refer to time-aligned translation tasks, including OMR and transcription. The advanced version puts forward the question of how data in one modality may help the training of LM in other modalities in general.}
    \label{fig:triangle-advanced}
\end{figure}

In the Grand Challenge Statement Section we presented a specific setup where data are abundantly available in A and V. Here we generalize the setup and present an advanced challenge setup (Fig.\ref{fig:triangle-advanced}). Given the same three elements as in Fig.\ref{fig:basic-setup}, we introduce several additional goals: 
\begin{itemize}
    \item Transcription from performance motion to score, inferring quantization, note grouping, and other visual decisions at the discretion of the musician that communicates intentions.
    \item Rendering audio from music score, inferring the performance techniques.
    \item Transcription from music audio to score, inferring quantization, note grouping, etc.  
    \item Improving $\mathrm{LM}^\mathrm{V}$ and $\mathrm{LM}^\mathrm{A}$ beyond uni-modal training potentials.
\end{itemize}

More generally, under the assumption of LMM, what is the proper treatment to the translation between all three modalities, A, V, and M, given an arbitrary set of available data? 

We hypothesize it may be beneficial to construct a hidden pan-modality core LM that facilitates operations in single modalities and across modalities. The core LM can be a hyper network that configures uni-modal models. 

\section{Conclusion}
In this grand challenge proposal, we present Language Model Mapping (LMM) in Multimodal Music Learning to explore fundamental connections between signals of different modalities. From an AI perspective, LMM serves as a path toward learning actions from both raw sensory inputs and abstract symbols, and in practice, it aims at human-like efficient learning. We introduce a basic challenge setup to specify a concrete case where LMM is required to solve a multimodal music learning problem. We then use more advanced cases to show that LMM in general is aimed at leveraging resources in one modality to bring meanings to model training in other related modalities. 






\bibliographystyle{iccc}
\bibliography{main}

\end{document}